\documentclass[12pt,preprint]{aastex}
\usepackage{amsmath,amssymb,graphicx,natbib}

\newcommand{\nz}{\hat{\mathbf{e}}_z}

\newsavebox{\astrutbox}
\sbox{\astrutbox}{\rule[-5pt]{0pt}{20pt}}

\newcommand{\vv}{{\mathbf v}}

\newcommand{\BB}{{\mathbf B}}

\newcommand{\gapprox}{\lower.4ex\hbox{$\;\buildrel >\over{\scriptstyle\sim}\;$}}
\newcommand{\lapprox}{\lower.4ex\hbox{$\;\buildrel <\over{\scriptstyle\sim}\;$}}

\slugcomment{To be submitted to ApJ Letters.} 
\shorttitle{Magetorotational turbulence in stratified boxes} 
\shortauthors{Bodo, Cattaneo, Mignone \& Rossi} 

\begin{document} 
 
\title{On the convergence of Magnetorotational turbulence  in stratified isothermal shearing boxes} 
 
 \author{G. Bodo\altaffilmark{1},        
         F. Cattaneo\altaffilmark{2}, 
         A. Mignone\altaffilmark{3},
         P. Rossi\altaffilmark{1} }
  
 \altaffiltext{1}{INAF, Osservatorio Astrofisico di Torino, Strada Osservatorio 20, 10025 Pino Torinese, Italy}
 
 \altaffiltext{2}{The Computation Institute, The University of Chicago, 
              5735 S. Ellis ave., Chicago IL 60637, USA}

 \altaffiltext{3}{Dipartimento di Fisica, Univesit\'a di Torino, via Pietro Giuria 1, 10125 Torino, Italy} 
 
\begin{abstract} 
We consider the problem of convergence in stratified isothermal shearing boxes with zero net magnetic flux. We present results with the highest resolution to-date--up to 200 grid-point per pressure scale height--that show no clear evidence of convergence.  Rather, the Maxwell stresses continue to decrease with increasing resolution. We propose some possible scenarios to explain the lack of convergence based on multi-layer dynamo systems. 
\end{abstract} 

\section{Introduction}
The magneto-rotational instability (MRI) and magneto-rotational turbulence (MRT) provide an elegant framework to study the origin of enhanced angular momentum transport in accretion discs. Much effort has been devoted to understanding the nonlinear development of the MRI and the processes that control the saturation amplitude of the instability, since, ultimately this controls the transport efficiency. Because of the difficulties inherent in approaching a strongly nonlinear problem analytically, much of the work on  MRT has relied on numerical simulations with all their attendant idealizations and approximations. By far the most popular is the shearing-box approximation in which the computational domain is restricted to a region of small radial extent at a large radius in the disc. Under reasonable assumptions this can be mapped into a Cartesian layer with shearing-periodic boundary conditions in the radial direction.  
Because the shearing-box approximation conserves vertical magnetic flux it is important to distinguish two types of simulations: those with finite initial (vertical) flux and those with zero initial flux. If the flux is finite there is a linear instability with a well defined growth rate and wavelength of maximum growth whose values are determined by the amount of flux  \citep{Balbus91}. In the nonlinear regime the amplitude of the Maxwell stresses--primarily responsible for angular momentum transport--is controlled by the amount of magnetic flux, and most crucially, remains finite in the ideal limit of vanishing dissipation. If, on the other hand, the initial flux is zero, the domain could in principle de-magnetize completely and relax to a state of uniform shear. If after a long time it does not, it must be because the magnetic field is being regenerated by turbulent motions. In this case, the MRI does not manifest itself as an exponentially growing linear instability, rather it is a subcritical dynamo process. In this case,  the spatial scales of the dominant magnetic structures and the efficiency of the angular momentum transport are determined by the dynamo itself. Two questions naturally arise: what kind of dynamo action can be sustained in a shearing-box, namely small-scale or large-scale, and what happens to the dynamo when the diffusivity, numerical or otherwise, becomes vanishingly small.  Addressing these issues has turned out to be a major and complex undertaking, even within the idealized framework of the shearing-box approximation. 

The first question is not specific to MRI driven  dynamos but to dynamos in general. Under what circumstances does a dynamo generate substantial amount of magnetic flux has been a long standing problem in astrophysical dynamo theory. Large-scale dynamos are often associated with flows lacking reflectional symmetry, or incorporating large scale shear, or a net flux of magnetic helicity through the boundaries, or any combination of the above. The second question was originally posed by \citet{Fromang07} within the framework of unstratified, homogeneous shearing boxes and it has since become known as the problem of convergence. Simply stated, a family of solutions of the MRI equations does {\it not} converge if the Maxwell stresses tend to zero as the dissipation tends to zero. Although, superficially, the convergence problem might seem mostly a matter of numerics, and indeed originally it was framed that way, actually it is not. Understanding why some shearing-box models converge and some do not is a fundamental question about nonlinear dynamo action in centrifugally stable systems. 
It is now commonly accepted that homogeneous, unstratified shearing boxes without explicit dissipation--these were the cases originally considered by \citet{Fromang07} do not converge \citep[for recent reviews, see ][]{Fromang13, Turner14}. The reason for the lack of convergence may be related to the small-scale nature of the dynamo operating in these systems, or to the lack of a characteristic outer scale or to a combination of these two factors \citep{Bodo11}.  All other cases are not as clear.

In the present paper we address the problem of convergence, or lack thereof, in the stratified isothermal case without explicit dissipation. This is the simplest shearing-box model with nontrivial stratification. Despite the  simplicity of the models, the dynamo that operates in these systems is far from simple. In an isothermal atmosphere with linear gravity reversing in the middle, hydrostatic balance gives rise to a density stratification with an approximately Gaussian profile and most of the mass concentrated near the mid-plane. A seemingly turbulent dynamo operates in this dense, central  region while propagating wavelike magnetic activity patterns are observed in the tenuous overlying layers \citep{Gressel10}.
A  resolution study by \citet{Davis10} with resolution up to 128 grid-points per scale height concluded that there was strong evidence for convergence. This led several authors to declare this case as settled in favor of convergence \citep{Gammie12, Fromang13, Turner14}. Here, we extend this study to 200 grid-points with a similar, but not identical, setup and numerics to that of \citet{Davis10} and find {\it no} evidence for convergence, at least up to these resolutions. Our conclusion is, therefore, that the problem of convergence for stratified, isothermal shearing-boxes is very much still an open issue.

\section{Formulation} \label{methods} 
We perform a convergence study for  a three-dimensional compressible, isothermal,  stratified shearing box \citep[for a description of the shearing box model see][]{Hawley95}.  The simulations start from a  layer in hydrostatic equilibrium. Assuming the vertical gravity of the form
$- \Omega^2 z$, where $\Omega$  is the orbital frequency and $z$ is the vertical coordinate, the density distribution takes the form
\begin{equation}
\rho = \rho_0 \exp(-z^2 / H^2),  \
\end{equation}
where $\rho_0$ is the value of density on the equatorial plane, $H$ is the scale height given by
 \begin{equation}
H = \frac{\sqrt{2} c_s}{\Omega},
\end{equation} 
and $c_s$ is the isothermal sound speed. Taking $1/\Omega$ as the unit of time, $H$ as the unit of length and $\rho_0$ as the unit of density,   the ideal MHD equations  for a keplerian shearing box, can be written in dimensionless form as

\begin{equation}
\frac{\partial \rho}{\partial t} + \nabla \cdot \left( \rho \vv \right) = 0,
\end{equation}

\begin{equation}
\frac{\partial \vv}{\partial t} + \vv \cdot \nabla \vv + 2  \nz \times \vv = \frac{\BB \cdot \nabla \BB}{\rho} - \frac{1}{\rho} \nabla 
\left(  \frac{\BB^2}{2} + P \right) - \nabla \left( -\frac{3}{2}  x^2  + \frac{1}{2}  z^2 \right)   ,
\end{equation}

\begin{equation}
\frac{\partial \BB}{\partial t} - \nabla \times \left( \vv \times \BB \right) = 0,
\end{equation}
where $\BB$, $\vv$, $\rho$ and $P$ denote, respectively,   non-dimensional magnetic  field intensity, velocity, density and pressure. In addition we assume an isothermal equation of state. Note that we absorbed a factor of $\sqrt{4 \pi}$ into the definition of $\BB$. 

 The simulation domain covers the region
 \begin{equation}
-0.5  < x < 0.5 , \qquad 0 < y < \pi , \qquad -3 < z < 3.
\end{equation}

The boundary conditions are periodic in $y$--the azimuthal direction, shear periodic in $x$--the radial direction, and impenetrable and stress-free in $z$. On the horizontal planes at $z = \pm 3$ we assume hydrostatic balance, and that the magnetic field is purely vertical. We note that these conditions allow a net flux of magnetic helicity through the boundaries unlike  those in \citet{Davis10} who adopt periodic conditions in $z$. Generically, it is found that in these stratified, isothermal models the ``vertical" boundary conditions have little effect on the qualitative structure of the solutions \citep{Davis10, Shi10, Oishi11, Gressel10}

We carried out a series of simulations at different resolutions with, respectively, 32, 64, 128 and 200 grid-points per scale height. The largest grid is, thus, $ 200 \times 600 \times 1200$. 
 Initially  the  magnetic field has the form
\begin{equation}
\BB = B_0 \sin \left( \frac{2 \pi x}{H} \right) \nz,
\end{equation}
where $B_0$ is chosen so as to give a ratio between thermal and magnetic pressure of $1600$. Clearly there is no net magnetic flux threading the box.  A small random perturbation  in the azimuthal component of the velocity is introduced to trigger the instability. 
The simulations were carried out  with  the PLUTO  code \citep{PLUTO} which allows the choice between several different  numerical schemes. For the present work we opted for third order accurate parabolic reconstruction, constrained transport method for the magnetic field evolution and HLLD Riemann solver \citep{Miyoshi05, Mignone07}.

\section{Results} \label{results} 
\begin{figure}[htbp]
   \centering
   \includegraphics[width=10cm]{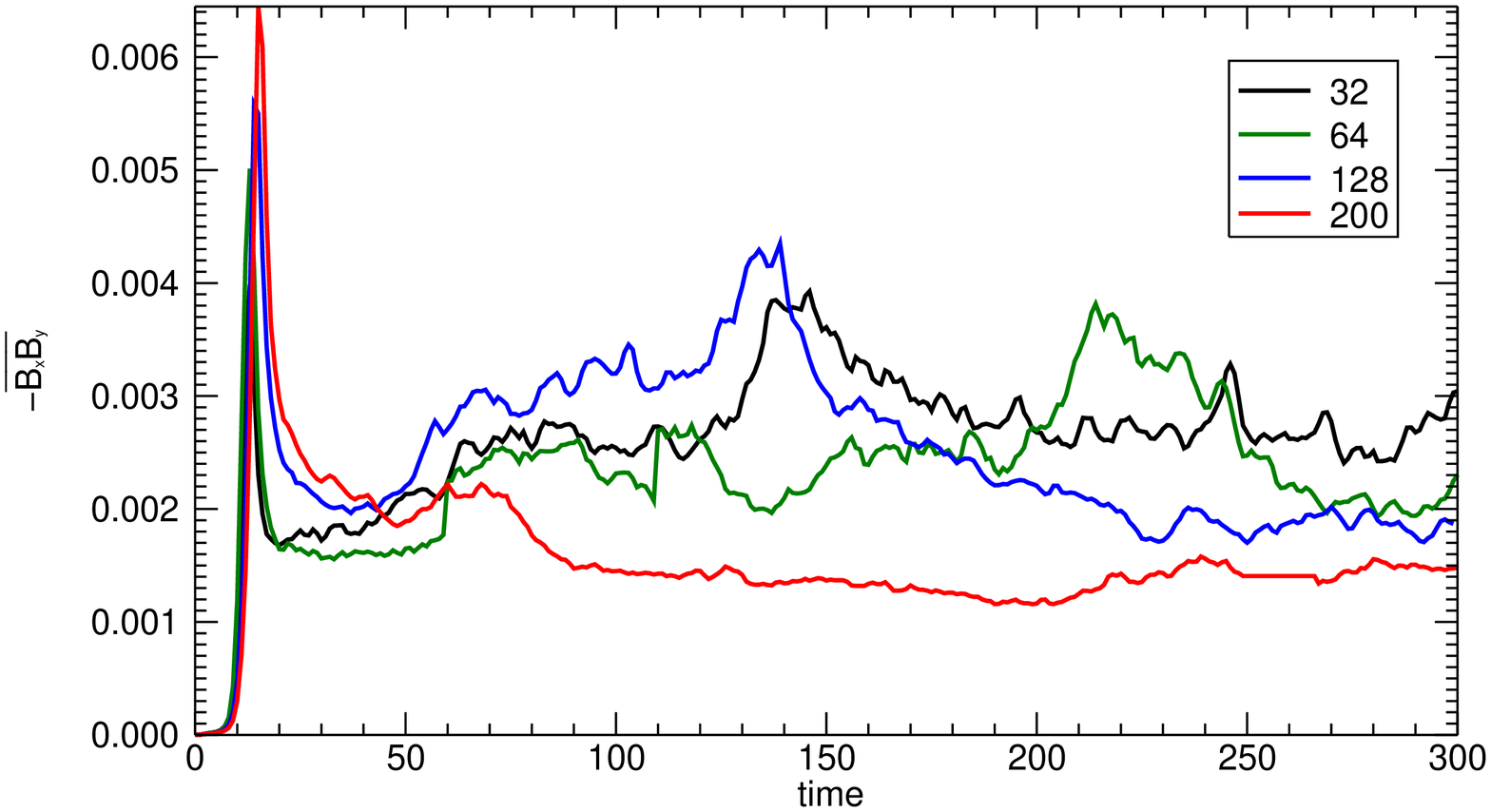} 
   \caption{Volume averaged Maxwell stresses as a function of time (measured in units of $1/\Omega$)}
   \label{fig:maxwtime}
\end{figure}

It is helpful to introduce the following notation: if $f$ is a generic function of space and time, we indicate a volume average by $\bar f$, an average over horizontal planes by $\tilde f$ and a time average by $\langle f \rangle$. The main result of our study is summarized in Fig. \ref{fig:maxwtime}, where we show the time history of the volume averaged Maxwell stresses, i.e. $  - \overline {B_xB_y}$,  for the four simulations with increasing resolution. As it is usual in these type of simulation, we observe that, after an initial transient that lasts about 50 units of time, the  stresses fluctuate around some average value. The amplitude of the fluctuations strongly decreases in the highest resolution simulation, similarly to what happens in the homogeneous periodic case \citep{Bodo11}. For the three simulations up to 128 points per scale height the value of Maxwell stresses seems to fluctuate around similar values and if we had limited ourselves to these results we would have reached the same conclusion as in \citet{Davis10}, i.e. that the stratified simulations seem to converge and give an efficiency of the transport independent from resolution. However, if we look at the curve corresponding to a resolution of 200 points per scale height, it  is systematically lower than the other ones. 

\begin{figure}[htbp]
   \centering
   \includegraphics[width=10cm]{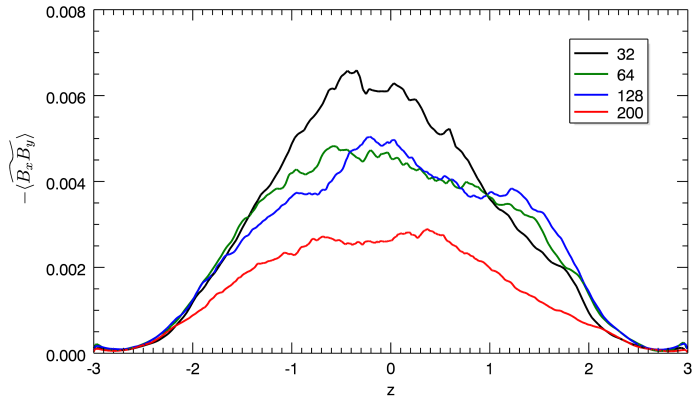} 
   \caption{Horizontal plane and time average of Maxwell stress as a function of $z$.}
   \label{fig:maxwz}
\end{figure}

We can get a more precise evaluation on how the efficiency of the transport changes with resolution  by  considering the horizontal plane time averaged (computed excluding the initial transient phase) Maxwell stresses, i.e. $ { \langle -\widetilde{B_xB_y} \rangle}$,  shown as a function of $z$ in Fig. \ref{fig:maxwz}. The stresses  decrease from the resolution of 32 points to 64 points, stay constant at 128 points and decrease again at the highest resolution of 200 points. In addition, looking at the behavior as a function of $z$, we see that the stresses are concentrated in a region around the equatorial plane and have a strong decrease in the high altitude regions.
   
\begin{figure}[htbp]
   \centering
   \includegraphics[width=10cm]{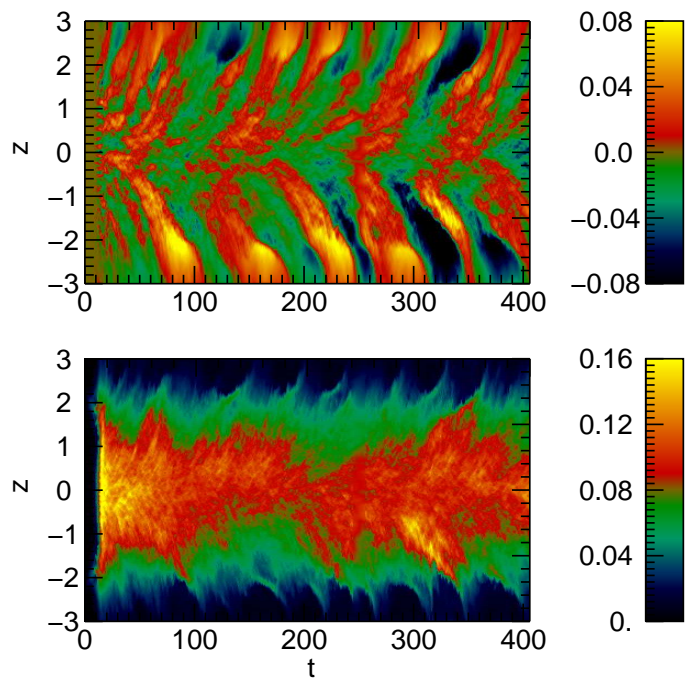} 
   \caption{The top panel of the figure shows the distribution of $\tilde B_y $ as a function of $t$ and $z$,  while the bottom panel shows the distribution of the r.m.s. value of the fluctuations  $\sqrt{\widetilde {{\delta B_y}^2}}$}
   \label{fig:butterfly}
\end{figure}

A major difference between the homogeneous, periodic case and the present one is the presence of an average magnetic field mostly in the toroidal direction. If we represent the distribution of $\tilde B_y$ as a function of $t$ and $z$, as we do in the top panel of Fig.  \ref{fig:butterfly}, we can observe cyclic patterns propagating away from the equatorial plane, that become more evident at high altitudes and have been observed  in all previous isothermal stratified simulations.   The bottom panel of the same figure shows the r.m.s. value of the fluctuations of $B_y$, i.e. $\sqrt{\widetilde {{\delta B_y}^2}}$, as a function of $t$ and $z$, where $\delta B_y$ is defined as
\begin{equation}
\delta B_y = B_y  -  \tilde B_y,
\end{equation}
indicating that the fluctuation level decreases  away from the equatorial region.

\begin{figure}[htbp]
   \centering
   \includegraphics[width=10cm]{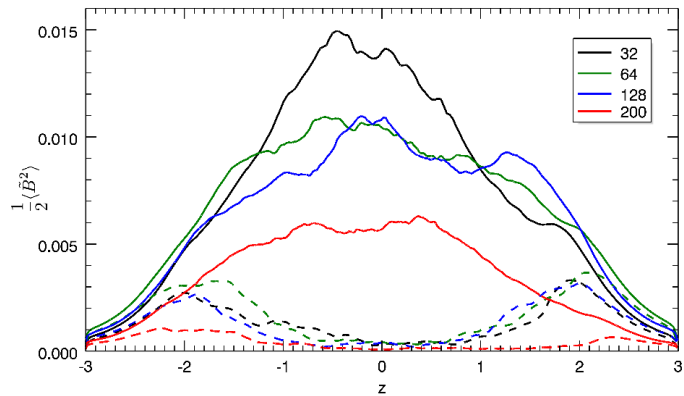} 
   \caption{Horizontal plane and time average of the total magnetic energy (solid lines) and of the magnetic energy of the mean field (dashed line) as a function of $z$}
   \label{fig:magen}
\end{figure}

These considerations can be made more quantitative by comparing the behavior of  the total magnetic energy  and the magnetic energy of the mean field as a function of height, which we do in Fig. \ref{fig:magen}.  The energy of the mean field is negligible in the equatorial region, therefore, in that region, the main contribution comes from fluctuations,  on the other hand it becomes comparable to that of the fluctuations or dominant in the high altitude regions.  In the highest  resolution case we have a clear decrease both in the fluctuations and the mean. 

\section{Conclusions} \label{conclusions}
We have revisited the problem of MRI driven turbulence in isothermal, stratified shearing boxes with zero net (vertical) magnetic flux and no explicit dissipation. We have extended our study to the highest resolution to date and find that, contrary to previously made claims based on lower resolution studies, the solutions do not converge, or at any rate, there is no convincing evidence of convergence. The average Maxwell stresses, principally responsible for the angular momentum transport, continue to decrease with increasing resolution. 

This conclusion can be further elaborated in terms of simple models of the types discussed by \citet{Blackman04} and \citet{ Gressel10} consisting of coupled dynamo systems operating in different regions.  One dynamo system is confined to the mid-plane where most of the mass is concentrated and gravity reverses, the other operating in the  tenuous overlying regions. The second dynamo is assumed to be of the mean-field type and be responsible for the generation of the magnetic structures that appear in the form of upward propagating dynamo waves.
As for the nature of the mid-plane dynamo system two possibilities readily come to mind.
One is that the motions in the mid-plane are driven by  small-scale  dynamo action similar to that observed in unstratified shearing-boxes. The justification for this assumption is that gravity is weak near the mid-plane. In this scenario the overlying mean-field dynamos are driven by the magneto-rotational turbulence in the mid-plane. The source of the turbulence is a subcritical dynamo instability. The other possibility is that the mean-field dynamos generate enough mean toroidal field in the mid-plane to drive an azimuthal MRI whose non-linear development drives the turbulence that, in turn, drives the mean-field dynamos. Although the outward manifestation of these two scenarios is the same, the reason for the apparent non-convergence is different. 

In the first scenario, the lack of convergence of the overall system follows from the non-convergence of the small-scale dynamo operating in the mid-plane, which can plausibly be reconstructed to the non-convergence of the unstratified homogeneous cases. If this analysis is the correct one, in the isothermal case, because  most of the mass is concentrated in a region where there is practically no gravity, stratification does not help to resolve the convergence problem.
It is useful to note that the above argument reduces the convergence issue for the stratified case to the convergence issue for the homogeneous case. In other words, the former does not converge because the latter does not converge. Contrariwise, it could be argued that if the homogeneous case were to converge so would the stratified one. At present, there is convincing numerical evidence from several different groups that in the absence of explicit dissipation the unstratified, homogeneous case does not converge
\citep{Fromang07, Pessah07,  Guan09, Simon09, Bodo11}. The case in which dissipation is included explicitly is not so clear cut  \citep[see, for instance, the comments in][]{Turner14}. Although it is often asserted that the homogeneous case with explicit dissipation converges \citep{Fromang13, Gammie12}, as far as we can tell, all the numerical evidence supporting these assertions originates from a single paper, namely that of \citet{Fromang10}. And although the simulations described therein remain an impressive numerical tour de force, we would argue that they are not such as to settle the issue of convergence unequivocally. We hope however that in the near future other attempt will be made to settle the issue of convergence in the presence of explicit dissipation conclusively.

In the second scenario the non-convergence derives from the inability of the mean-field dynamo to operate at high magnetic Reynolds numbers--be they real or numerical. This is a well known effect that has received much attention and goes back to the original works by  \citet{Cattaneo92, Kulsrud92} and \citet{Gruzinov94}. In this case as the dissipation decreases so does the generated mean toroidal field needed to destabilize the azimuthal MRI in the mid-plane regions. Eventually the mean toroidal field is so weak that the system become indistinguishable from the one in the first scenario with all its attendant limitations. Should  this analysis turn out to be correct it is interesting that it applies to a case with boundary conditions that allow a net flux of magnetic helicity.

Undoubtedly other scenarios can be constructed that agree with the numerical evidence, provide an explanation for the non-convergence and highlight the role of other physical processes. However, to quote from one of the authors' favorite poems ``whatever the reason his heart or his shoes"  \citep{Seuss57} the stratified isothermal shearing-box appears not to converge.

\section{Acknowledgment}
The authors would like to thank the anonymous referee for the careful reading of the manuscript and for several valuable comments.
This work was supported in part  by the National Science Foundation 
sponsored Center for Magnetic Self Organization at the University of Chicago. 
Computations were performed at the J\"ulich Supercomputing Center thanks to a DEISA
DECI grant.


\end{document}